\documentclass{elsart}

\usepackage{amssymb}

\usepackage{graphicx}
\usepackage{dcolumn}

\begin{document}

\begin{frontmatter}


\title{A Monte Carlo method to generate fluorescence light in
  extensive air showers} 

\author{Vitor de Souza\corauthref{cor1}}
\author{, Henrique M. J. Barbosa}
\author{and Carola Dobrigkeit}
\corauth[cor1]{vitor@ifi.unicamp.br}
\address{
Instituto de F\'{\i}sica Gleb Wataghin, Universidade Estadual de Campinas\\
13083-970 Campinas-SP, Brasil }

\begin{abstract}
A new approach to simulate fluorescence photons produced in extensive
air showers is described. A Monte Carlo program based on CORSIKA
produces the fluorescence photons for each charged particle in the
development of the shower. This method results in a full
three-dimensional simulation of the particles in a shower and of the
fluorescence light generated in the atmosphere. The photons produced
by this program are tracked down to a telescope and the simulation of
a particular detector is applied. The differences between this method
and one-dimensional approaches are quantitatively determined as a
function of the distance between the telescope and the shower core.
As a particular application, the maximum distance at which the Auger
fluorescence telescope might be able to measure the lateral
distribution of a shower is determined as a function of energy.
\end{abstract}

\end{frontmatter}

\section{\label{introduction} Introduction}

There are several questions yet to be solved about cosmic rays with
energy above $10^{18} \ eV$. The most exciting topics concern the
existence of a cut off in the energy spectrum, the possible anisotropy
of the sources, the chemical composition and the propagation
mechanisms.

Particles with these energies arrive on Earth with a very low flux and
therefore can only be studied by the detection of the extensive air
showers (EAS) they produce in the atmosphere. The present HiRes
\cite{bib:hires} and the AGASA \cite{bib:agasa} experiments have
contributed to our understanding of the nature of high energy cosmic
rays and future detectors (Auger
\cite{bib:auger-nim}, EUSO \cite{bib:euso-artigo}, OWL
\cite{bib:owl-artigo} and  Telescope Array \cite{bib:telescope:array}) are
very promising due to the increase in collection area and the use of
new technologies.

The fluorescence technique has been used successfully by the HiRes
experiment and is also going to be used by the new generation of
detectors: Auger, EUSO, Telescope Array and OWL. Fluorescence
detectors measure the EAS by counting, as a function of depth, the
photons produced by the excitation of nitrogen molecules due to the
passage of the charged particles in the shower. The determination of
the longitudinal profile and the time the photons have hit the
detector allow the reconstruction of the energy and direction of the
primary particle.

Monte Carlo simulations have been widely used by cosmic ray physicists
to optimize their detectors and to determine reconstruction
methods. One of the most popular and tested programs known in the
cosmic ray community is CORSIKA \cite{bib:corsika}. CORSIKA simulates
the development of the particles in a shower in a high level of
sophistication including three-dimensional particle transport,
electromagnetic interactions, decay of unstable particles, multiple
scattering, deflection in the Earth's magnetic field and several high
energy hadronic interaction models. Comparisons of CORSIKA simulations
and experimental data can be found in \cite{bib:corsika-teste}.

A shower with energy $10^{19} \ eV$ has more than $10^{10}$ particles
and a Monte Carlo simulation would take a long time to follow all of
them. Hillas \cite{bib:thin} has overcome this problem by using a
thinning algorithm in Monte Carlo simulations. This procedure follows
only a fraction of the particles with energy below a certain limit and
conserve the total energy in the shower. The use of this method has
reduced the computational time to practical limits without decreasing
significantly the quality of the simulations.

In contrast to the simulation of particles in a shower, the simulation
of the fluorescence light produced by the passage of these particles
in the air has been done with almost no detail. Calculations made so
far were done based on one-dimensional simulations of the fluorescence
photons following an average function given by a Gaisser-Hillas
\cite{bib:gaisser-hillas} profile. Apparently, the crude treatment
given to the simulation of the fluorescence component was justified by
the optics and electronics of the present generation of
telescopes. However, for the future detectors this justification is
certainly not going to be valid.

In reference \cite{bib:sommers_size}, Sommers has made analytic
calculations to show the possibility of detecting the lateral
distribution of a shower with a fluorescence detector. His ideas have
been exploited by G\'ora et al. \cite{bib:gora_size} using simulations
based on a Gaisser-Hillas longitudinal profile and a NKG
\cite{bib:nkg,bib:nkg_crsk} lateral distribution function.

More recently, other groups have announced new fast simulations in
development \cite{bib:seneca,bib:casc1d} which may be used to produce
fluorescence photons. These programs are based on a combination of
numerical and Monte Carlo approaches and special attention must be
taken in order not to reduce the intrinsic shower fluctuations.

The work presented here explains how we generate fluorescence photons
in the CORSIKA framework. Our approach allows the simulation of this
component of the shower with great detail. For each charged particle
produced in the shower development, the corresponding fluorescence
light is produced from the ionization energy loss and the are tracked
to the detectors. Atmospheric attenuation may be applied and the
particles continue to be followed by CORSIKA down to the observational
level resulting in a full hybrid (fluorescence + particles)
simulation. In particular, this simulation can be useful for the Auger
Collaboration due to its hybrid design.

The atmospheric attenuation is done by a simple implementation that
simulates the extinction of photons according to particular
atmospheric properties. The user is allowed to implement different
atmospheric extinction coefficients. Excluding the showers simulated
in section \ref{sec:certification} where no attenuations has been
applied, all simulations in this paper were done with a midlatitude US
Standard Atmosphere and scattering of light by aerosols and molecules
was taken into account. The aerosol horizontal attenuation length and
vertical scale height were set to $12 \ km$ and $1.2 \ km$,
respectively.  The mean free path for Rayleigh scattering was set to
$2974 \ g/cm^2$.  Scattering of the fluorescence photons is not
simulated in the present version which could be an important effect
for some specific geometries of detection and therefore should be
accounted for in future implementations.
  
Historically, the main advantage of using Monte Carlo simulations
comparing to analytic models has been considered to be the possibility
to reproduce the event by event fluctuations. Nevertheless, the
development of the softwares has extended the differences beyond that,
among which, we stress the possibility to inject almost all kind of
primary particles, the detailed description of the atmosphere and the
treatment of each shower component (hadronic and electromagnetic) in
an independent way with parameters updated from the accelerator
experiments.
     
Besides that, the simulation we has developed have some other
improvements if compared to analytic or one-dimensional models
regarding the fluorescence photons production. In a deeper level of
analysis, the approach we present here is able to simulate the
particle by particle fluctuation in the production of the fluorescence
photons, since each charged particles is treated independently.

Moreover, the time information is available for each single photon
taking into account its exact position in the atmosphere. This feature
would allow for the first time a detailed study of the snapshots
proposed in \cite{bib:sommers_size} with a good raytracing resolution.

Recently proposed experiments \cite{bib:fy:campinas,bib:fy:airfly} are
planning to measure the fluorescence 
yield in  more details than done so far by Kakimoto et al.
\cite{bib:kakimoto}. The fluorescence yield varies with the particle
energy and may be completely different for each particle type. The
simulation presented here can easily take into account the
dependencies of the yield as a function of energy and particle type
for each particle in the shower after they have been measured by the
proposed experiments. This great level of detail would be practically
impossible to be accounted for in a analytical model or in an
one-dimensional Gaisser-Hillas approach.

In this paper, we describe the simulation, test it against old and
simple approximations and illustrate the use of the program by
calculating the lateral size of the shower as a function of the
distance to a ground based telescope. This application shows that the
lateral distribution of showers with energies $10^{18}$, $10^{19}$ and
$10^{20} \ eV$ measured by detectors similar to the Auger telescopes
configuration can not be neglected when distances from the detector to
the shower axis are shorter than $4 \ km$, $8 \ km$ and $14 \ km$,
respectively.
 
\section{Fluorescence Light Simulation}
\label{sec:simulation}

We used CORSIKA as the basic framework for the generation of particles
in the shower. The detailed simulation implemented in CORSIKA makes
all the information needed to generate fluorescence photons for every
charged particle in a shower available. Explicitly, the position of
creation and death of the current particle and the energy deposit of
this particle in the atmosphere are accessible.

For each charged particle produced inside CORSIKA, a new subroutine
that generates the fluorescence light is called.  With the initial and
final positions $(\vec{r}_i, \vec{r}_f)$ and energies $(E_i,E_f)$ of
the particle the subroutine does a Monte Carlo simulation for the
production and propagation of fluorescence photons.

However, when the energy of a particle falls below a certain value, it
is neglected by CORSIKA and the information needed to create
fluorescence photons is not available. For electrons and positrons
below the energy cut, $50 \ keV$, we made the hypothesis that they
deposit all the available energy as they are discarded by CORSIKA.  We
also assume that the track length traveled by these particles ($\Delta
X$) is given by the Bethe-Bloch theory and the stopping range is
calculated assuming the continuous-slowing-down-approximation.

\begin{figure}[p]
\centerline{\includegraphics[width=13cm]{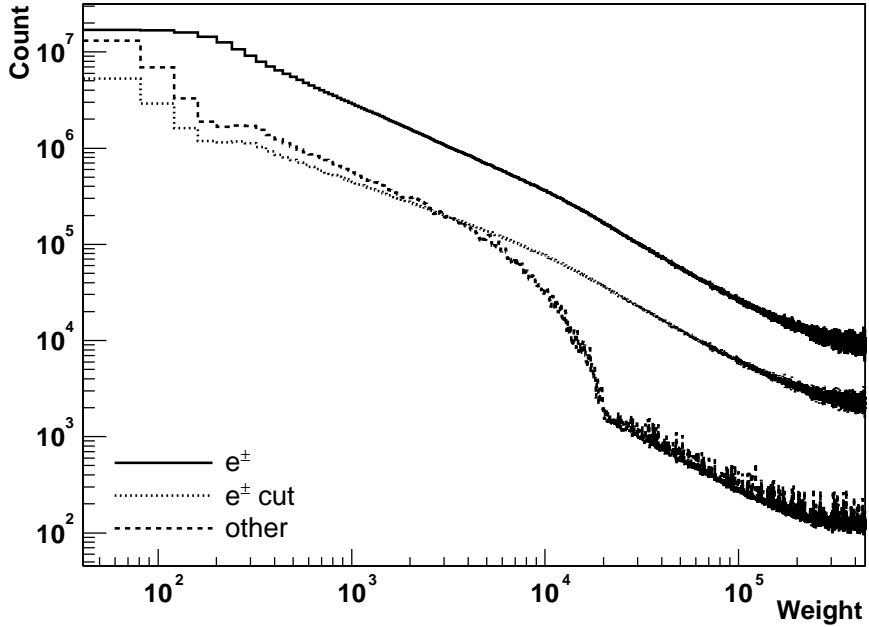}}
\caption{Mean distribution of weights for 10 proton showers of
  $10^{19}\ eV$
and zenith angle $45^o$. The thinning level was set to $10^{-6}$.}
\label{fig:histo_peso}
\end{figure}

As explained previously, in simulations of high energy EAS one usually
adopts a thinning method in order to reduce the number of particles to
be followed and stored. In this procedure a weight \emph{W} is
attributed to each particle. As can be seen in Figure
\ref{fig:histo_peso}, where we show the average weight of particles in
a $10^{19}\ eV$ shower simulated with thinning factor $10^{-6}$, many
particles have weight well beyond 1000.

In order to avoid the necessity of applying an unthinning procedure in
the detector simulation we considered the energy deposited by a single
particle as given by: $\Delta E = W \cdot
\Delta E^{single}_{particle}$. 

Since the total path ($\Delta X$) and the energy deposited by
particles ($\Delta E$) with energy above the energy cuts and by
electrons and positrons with energy below the cutoff are now known,
the number of photons produced by each of these particles can be
calculated using the fluorescence yield given by Kakimoto et
al. \cite{bib:kakimoto}:

\begin{equation}
N_{pht}  = \int \frac{dE}{dX} \; \rho \; dx \left(
\frac{A_1}{1 + \rho B_1 \sqrt{T}} + \frac{A_2}{1 + \rho B_2 \sqrt{T}} \right)
\left( \frac{dE}{dX} \right)_{1.4 \ MeV}^{-1} 
\label{eq:kakimoto}
\end{equation}

\noindent where the temperature $T$ and the density $\rho$ are functions
of the altitude and the constants are $A_1 = 89.0 \ m^2 kg^{-1}$, $A_2
= 55.0 \ m^2 kg^{-1}$, $B_1 = 1.8 \ m^3 kg^{-1} K^{-1/2}$, $B_2 = 6.5
\ m^3 kg^{-1} K^{-1/2}$ and $(dE/dX)_{1.4 \ MeV} = 1.8 \ MeV/(g/cm^2)$.

Finally, since fluorescence emission is isotropic, only a small
fraction of these photons is emitted towards the detectors and this
fraction depends only on the distance to the detector and on its
geometry.  Once the user has defined the position and radius of the
detectors for which the simulation is done, the average number of
photons emitted by a charged particle in the direction of the
detectors can be calculated as:

\begin{equation}
\overline{\aleph} =  N_{pht} \cdot \frac{d\Omega_{tot}}{4\pi}
\label{eq:ndet}
\end{equation}

\noindent where $d\Omega_{tot}$ is the solid angle subtended by 
all the detectors.

\begin{figure}[p]
\centerline{\includegraphics[width=13cm]{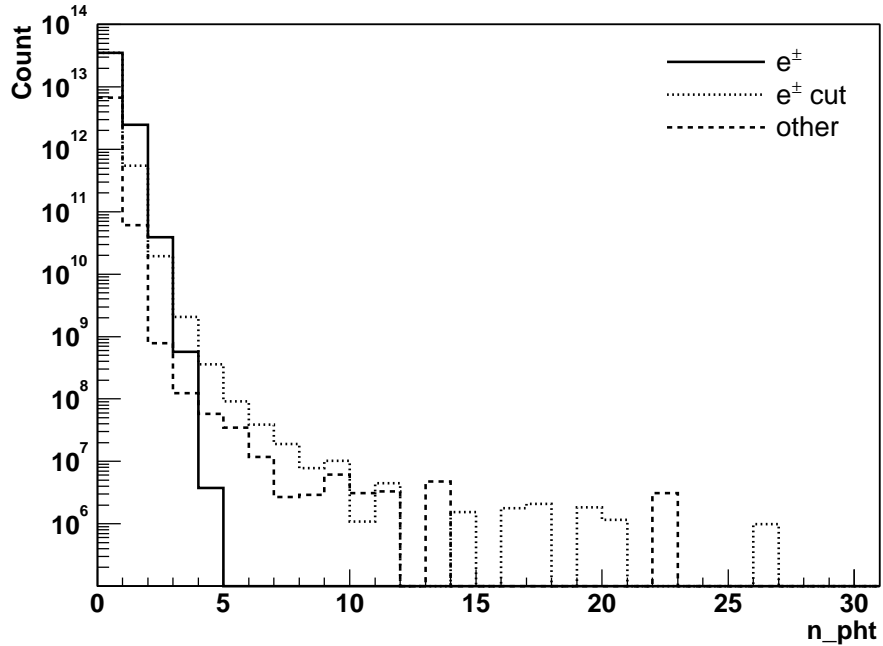}}
\caption{Distribution of the number of photons emitted by a single
particle toward the telescopes. Same events as in figure \ref{fig:histo_peso}.
The telescope was positioned $15 \ km$ away from the shower core.}
\label{fig:histo_npht}
\end{figure}

In general, we have $\overline{\aleph} < 1$ because the solid angle
subtended by the detectors is small compared to $4\pi$ and despite the
great amount of photons emitted, the probability that any of them will
be directed inside $d\Omega_{tot}$ is very small.  According to the
physical process involved, the final number of photons (\emph{N})
emitted towards the detectors, by the passage of one given particle,
can be described by a Poisson distribution with average
$\overline{\aleph}$. Figure \ref{fig:histo_npht} shows the
distribution of the number of photons emitted towards the detector,
for the same events as in figure \ref{fig:histo_peso}.

Knowing the number of photons to be emitted by the particle, its
trajectory is divided in \emph{N} equal intervals and one photon is
emitted from the center of each interval.  The emission point is set
to the middle of each interval and the direction of emission of the
photon is drawn inside the solid angle of one of the detectors.
Therefore, we can evaluate whether or not the photon's trajectory lies
within the solid angle subtended by the disk that defines the
detector.  If the intersection of the trajectory with the disk exists,
the emission position, propagation direction, time and wavelength of
photons that hit a detector are saved in a file. The wavelength is
drawn from the known fluorescence spectrum \cite{bib:bunner}.

\section{Tests}
\label{sec:tests}

The longitudinal profile of charged particles is the most important
feature of a cosmic ray shower measured by fluorescence detectors. The
longitudinal development of the shower is used to estimate the energy
and gives a hint on the mass of the primary particle.

In order to test the fluorescence photon production we have compared
the number of photons produced along the shower development and the
arrival time of these photons at a telescope simulated by our program
with an analytic calculation based on the longitudinal energy
deposited in the atmosphere as given by standard CORSIKA.

We have also compared the longitudinal profile of a real event
detected by the Fly's Eye telescope \cite{bib:fly51joule} with the
simulation of the same shower.

\subsection{Certification of the Number and Arrival Time of Photons}
\label{sec:certification}
 
We have simulated vertical showers with energy $10^{18} \ eV$.  For
this application, electrons were chosen as primary particle of the
showers because we would like to avoid confusions derived from the
energy channeled into neutrinos, high energy muons and nuclear
excitation in hadronic showers \cite{bib:song}.  The thinning factor
was set to $10^{-6}$ to reduce fluctuations in the longitudinal
profile and no atmospheric attenuation was applied.

\begin{figure}[p]
\centerline{\includegraphics[width=13cm]{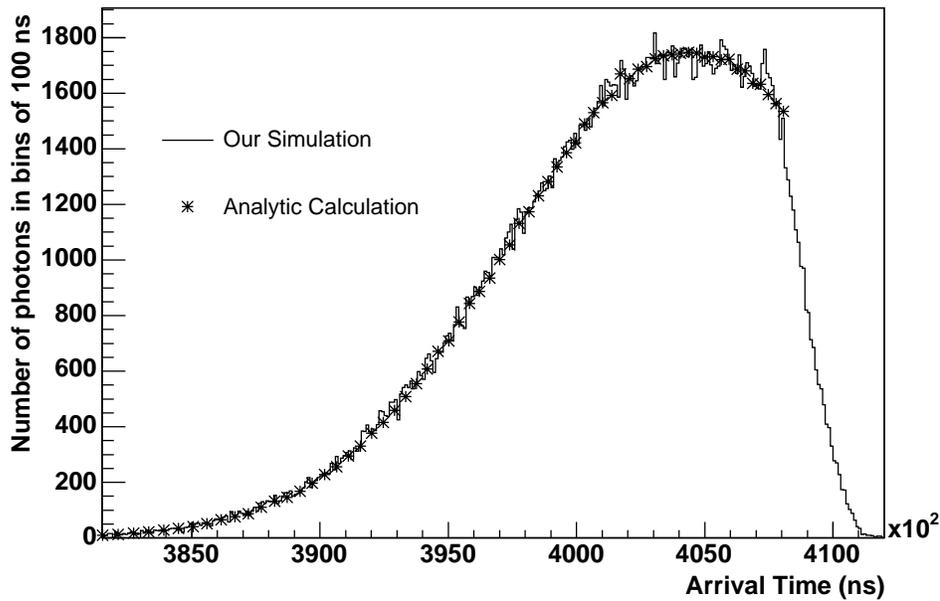}}
\caption{Comparison between the number of photons collected in bins of
  100 ns in the detector as a function of time as simulated by our program
  and determined from analytic calculations. Time starts when first
  interaction happened.} 
\label{fig:time}
\end{figure}

\begin{figure}[p]
  \centerline{\includegraphics[width=13cm]{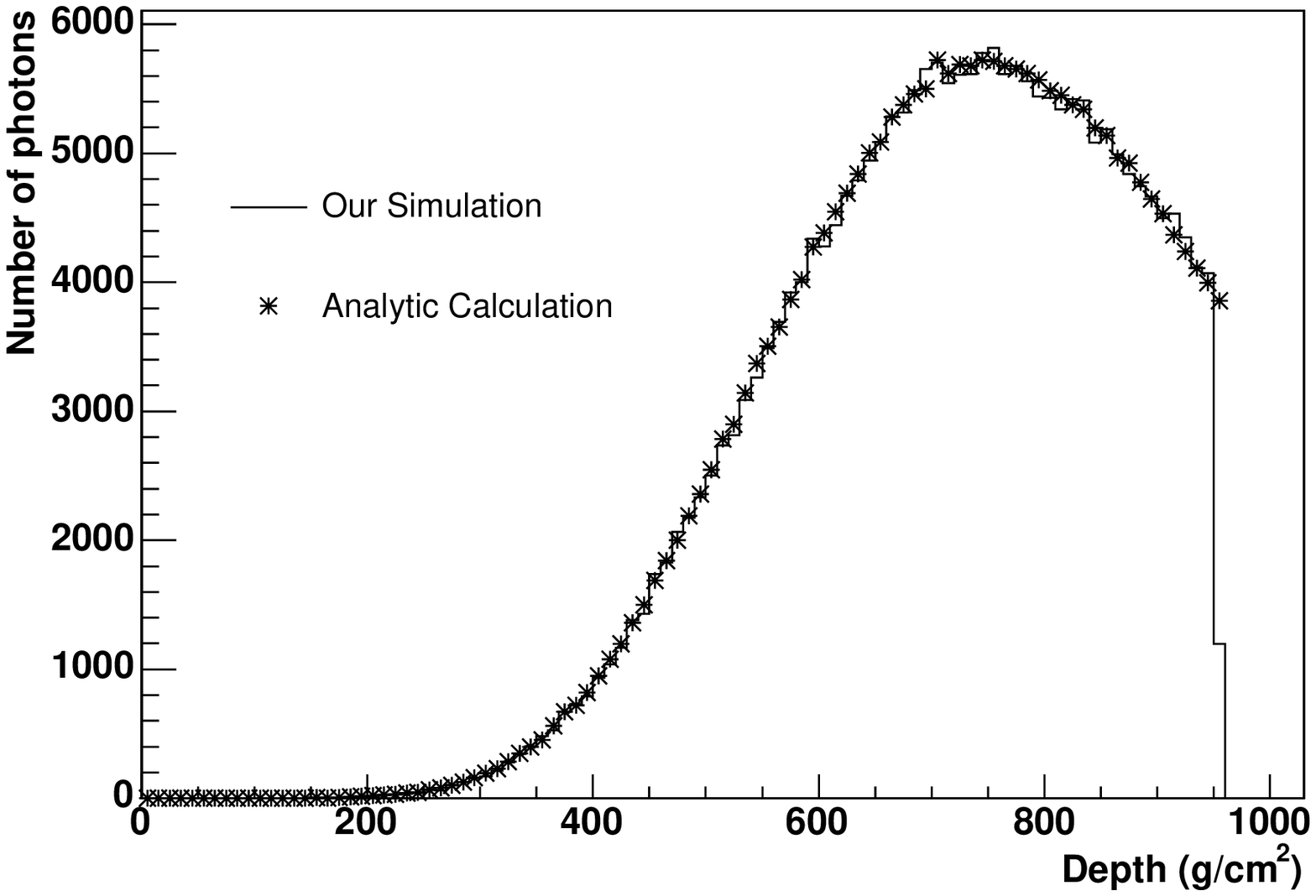}}
\caption{Comparison between the number of photons as a function of
  atmospheric depth calculated from the
  longitudinal energy deposited in the atmosphere and simulated by our
  program. The thinning factor as set to $10^{-6}$.} 
\label{fig:perfil}
\end{figure}

The fluorescence photons produced by our code are propagated down to a
telescope of radius $10 \ m$, positioned $40 \ km$ away from the
shower core.  For the studies presented in this section, we have
stored the height of production and the arrival time of the photons
detected by the telescope as can be seen in figures
\ref{fig:perfil} and \ref{fig:time}.  

The standard version of CORSIKA records the energy deposited in the
atmosphere by all particles in intervals of a given grammage.  We have
set the bin size to be 10 $g/cm^2$ and from the longitudinal energy
deposit profile we were able to calculate analytically the number of
photons produced in a bin and consequently the average arrival time of
this photons in the telescope.

To the amount of energy deposited in one bin corresponds a certain
amount of photons detected in the telescope that is given by equations
\ref{eq:kakimoto} and \ref{eq:ndet} where density and temperature can
be approximated to the values at the middle of the bin.  Since the
distance between the telescope and the middle of the bin along the
axis of the shower is known we can also calculate the time these
photons would hit the detector.

In this calculation all particles are considered to be emitted along
the axis of the shower so that to be able to compare with our
three-dimensional program we have positioned the telescope far away of
the axis ($40\ km$) in order to minimize the differences. However, the
number of photons measured by a regular fluorescence detector with few
meters of aperture would be very small at this distance.  To avoid
fluctuations due to low statistics we simulated a detector with $10
\ m$ of aperture. No telescope optics or electronics were simulated. 

Figure \ref{fig:perfil} shows, for a single shower, the longitudinal
profile of photons that end up to be detected in the telescope and the
results of the analytic calculation based on the energy deposited in
the atmosphere. The agreement between our simulation and the simple
calculations is perfect. Similar results are obtained for all showers
at different energies and inclinations.

Figure \ref{fig:time} shows the arrival time in bins of 100 ns as
given by our program and the analytic calculation. Again the agreement
is very good and similar results are obtained for all showers at
different energies and inclinations.

The comparisons shown in these graphics proves that our simulation
program is able to reproduce the right number and arrival time of
photons at the telescope as expected by simple arguments.

\subsection{Comparison to a Real Event}
\label{sec:real}

In 1995, the Fly's Eye experiment detected a very clear shower with
energy well beyond the GZK cut-off. This event has been published
\cite{bib:fly51joule} and we used the reconstructed longitudinal
profile as a test of the consistency for our program.

The famous shower has landed $13.6 \ km$ away from the telescope and
its primary energy and zenith angle were estimated as $320 \ EeV$ and
$43.9^o$, respectively. See \cite{bib:fly51joule} for details of the
reconstruction. We have simulated with our program 10 events with the
same energy and geometry of the Fly's Eye shower: 5 initiated by
proton, and 5 by iron.  The produced photons were propagated through
the optics of a telescope similar to the Auger Fluorescence Telescopes
\cite{bib:auger-nim}. The design of these telescopes is a Schmidt
camera with a corrector ring determining an aperture of radius $1.1 \
m$ and a spherical mirror with $30^o \times 30^o$ field of view
\cite{bib:lente}. The photomultiplier camera is formed by 440 
phototubes in a grid of $22\times20$ with $1.5^o$ field of view
each. In the simulations of the signal detected by the
photomultipliers we have added a poissonian noise with average equal
to the night sky background.

\begin{figure}[p]
\centerline{\includegraphics[width=13cm]{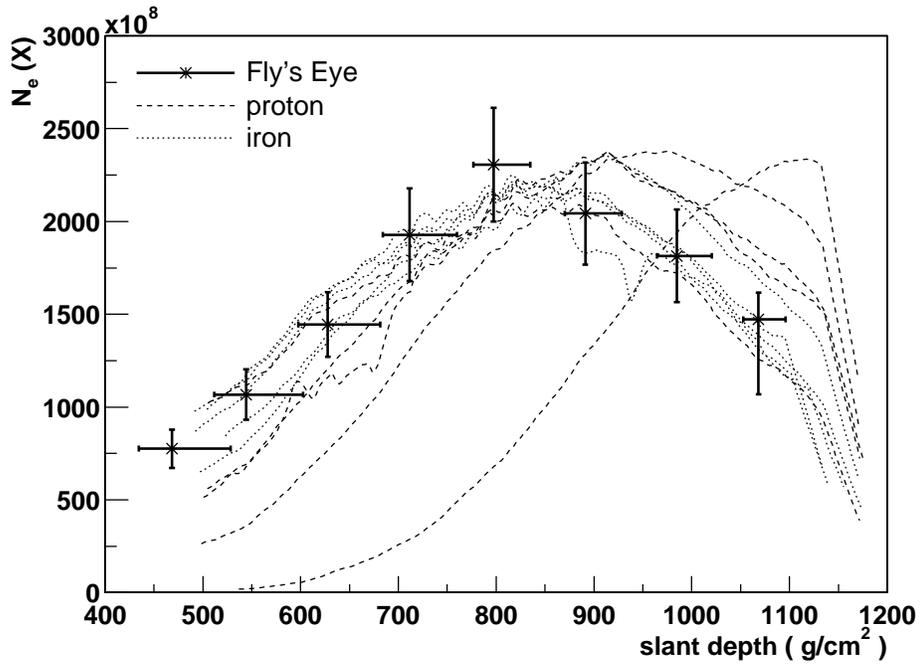}}
\caption{Comparison of the longitudinal profiles of five proton and five iron showers simulated by our
  code to the profile reconstructed by the Fly's Eye Collaboration.}
\label{fig:flydata}
\end{figure}

The simulation was reconstructed to give the longitudinal profile
according to the standard procedure described in
\cite{bib:fly51joule}.  Figure \ref{fig:flydata} shows a comparison of
the longitudinal profile determined by the Fly's Eye group with the
longitudinal profile simulated with our code. An integration of the
profile leads to the energy of the primary particle. The estimated
energy from the simulations is $3.4\pm.51(rms)\times 10^{20} \ eV$
which is in good agreement with the value estimated by the Fly's Eye
Collaboration.

\section{Applications}
\label{sec:application}

Several new applications are possible with this three dimensional
fluorescence and particle simulation. Among them, we could stress the
verification of hybrid reconstruction methods, the analysis of arrival
time of the photons in a telescope as suggested by Sommers
\cite{bib:sommers_size}, the determination of the ``missing energy''
\cite{bib:song} in a shower and the configuration of the new
generation of fluorescence detectors.

However, the first demanding application is to determine the
capability of a fluorescence detector to distinguish a one-dimensional
shower from a three-dimensional one. In other words, in which
conditions a fluorescence detector is going to detect the lateral
distribution of particles in a shower?

To answer this question, we have simulated two sets of 100 vertical
showers initiated by $10^{19}\ eV$ protons. The thinning factor was
set to $10^{-6}$. One set was simulated with the three-dimensional
program described here, and the other with an one-dimensional
simulation following a Gaisser-Hillas distribution. Ten telescopes
were positioned in a line from $2 \ km$ up to $20 \ km$ away from the
core of the shower, equally spaced in steps of $2 \ km$.

We have simulated the fluorescence detectors used by the Auger
collaboration described in section \ref{sec:real} and the quantitative
conclusions shown here might be very different for detectors with
different configurations, in particular for telescopes with different
pixel size.

\begin{figure*}[p]
\centerline{
\includegraphics[width=6.5cm]{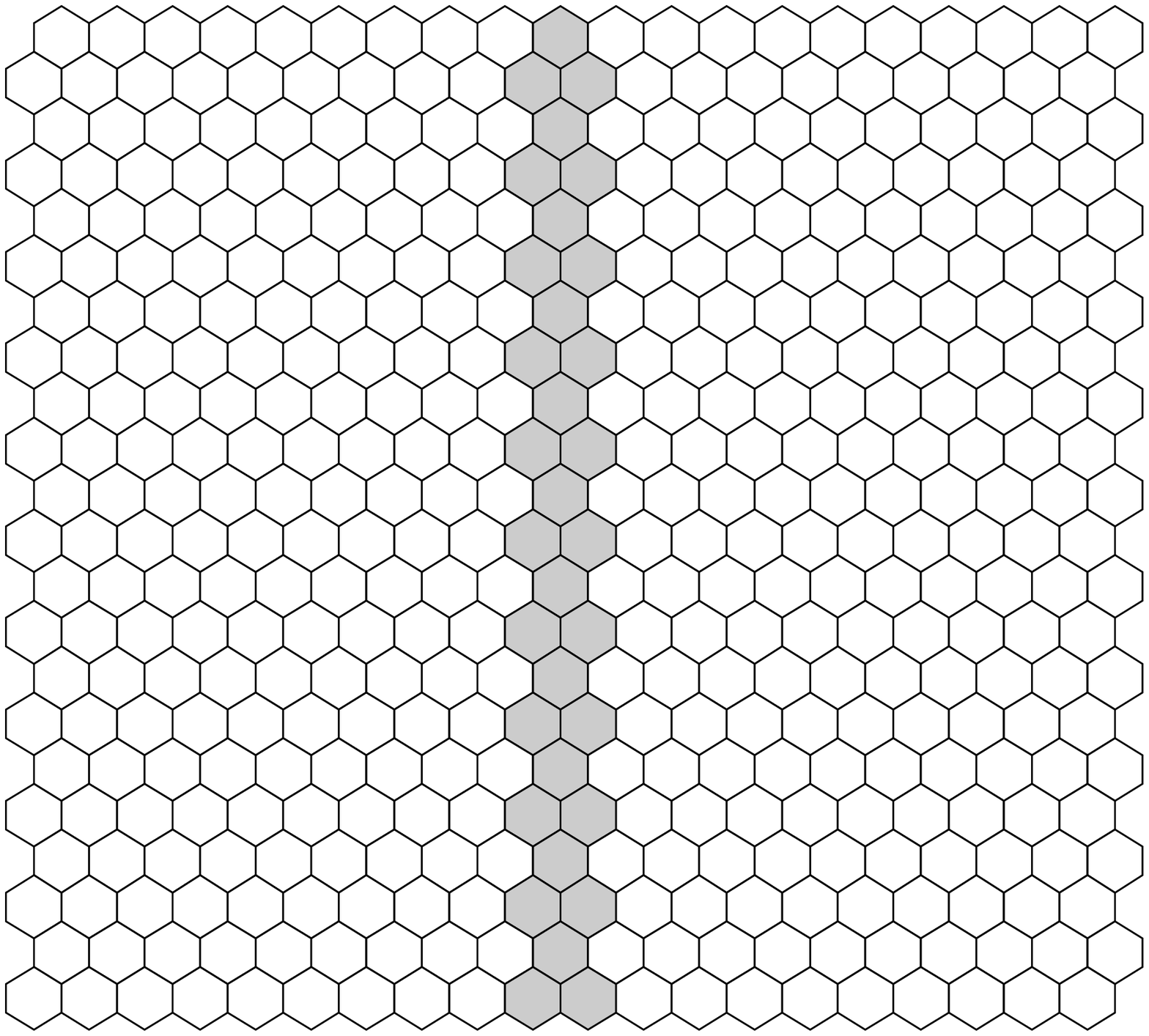}\hspace{0.8cm}
\includegraphics[width=6.5cm]{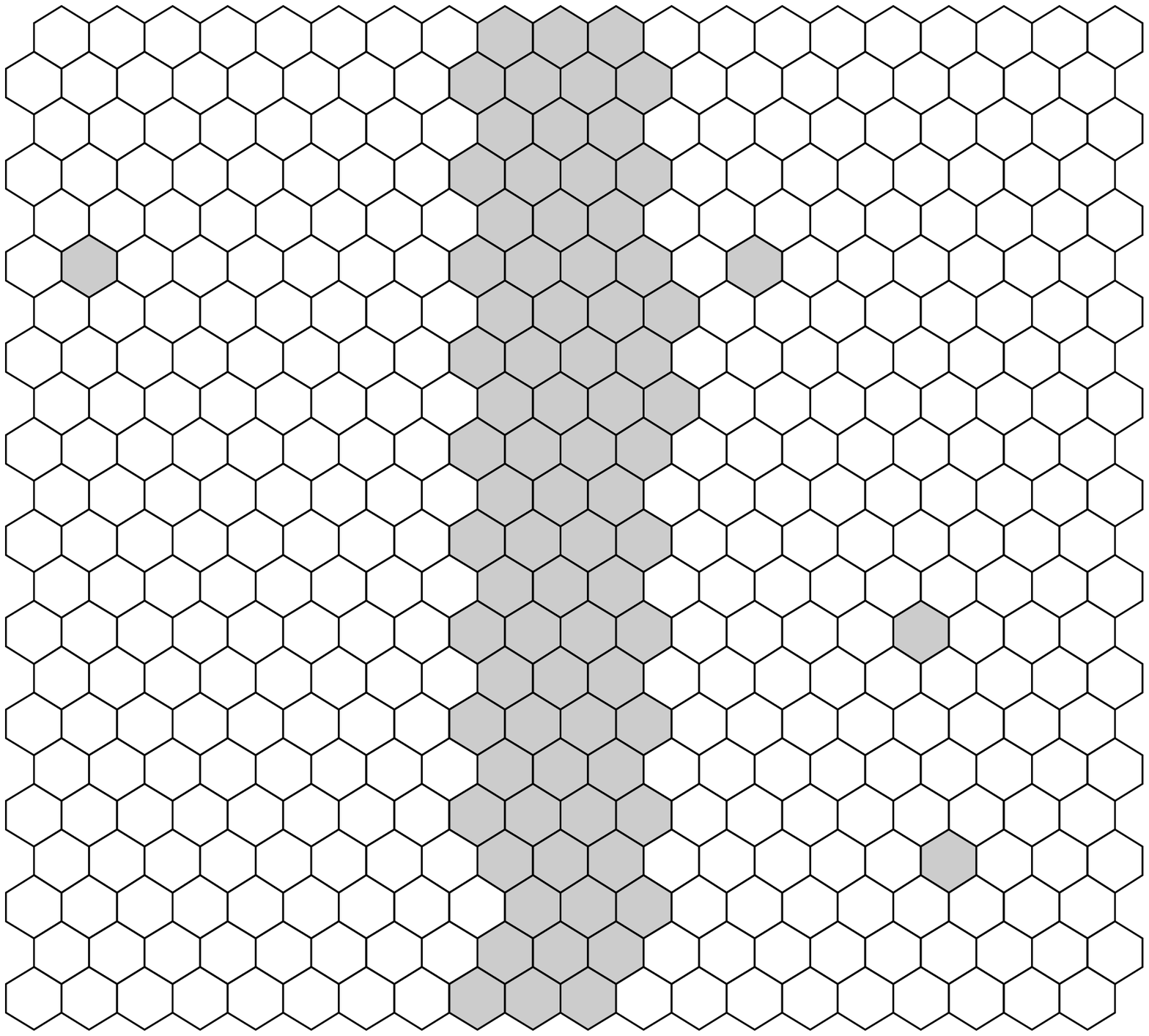}}
\caption{Distribution of the triggered pixels on the camera of a telescope
positioned $4\ km$ from the shower axis. 
The results of the one-dimensional and three-dimensional simulations
of the shower are shown
in the left and right panels, respectively.}
\label{fig:cam}
\end{figure*}

\begin{figure}[p]
\centerline{\includegraphics[width=13cm]{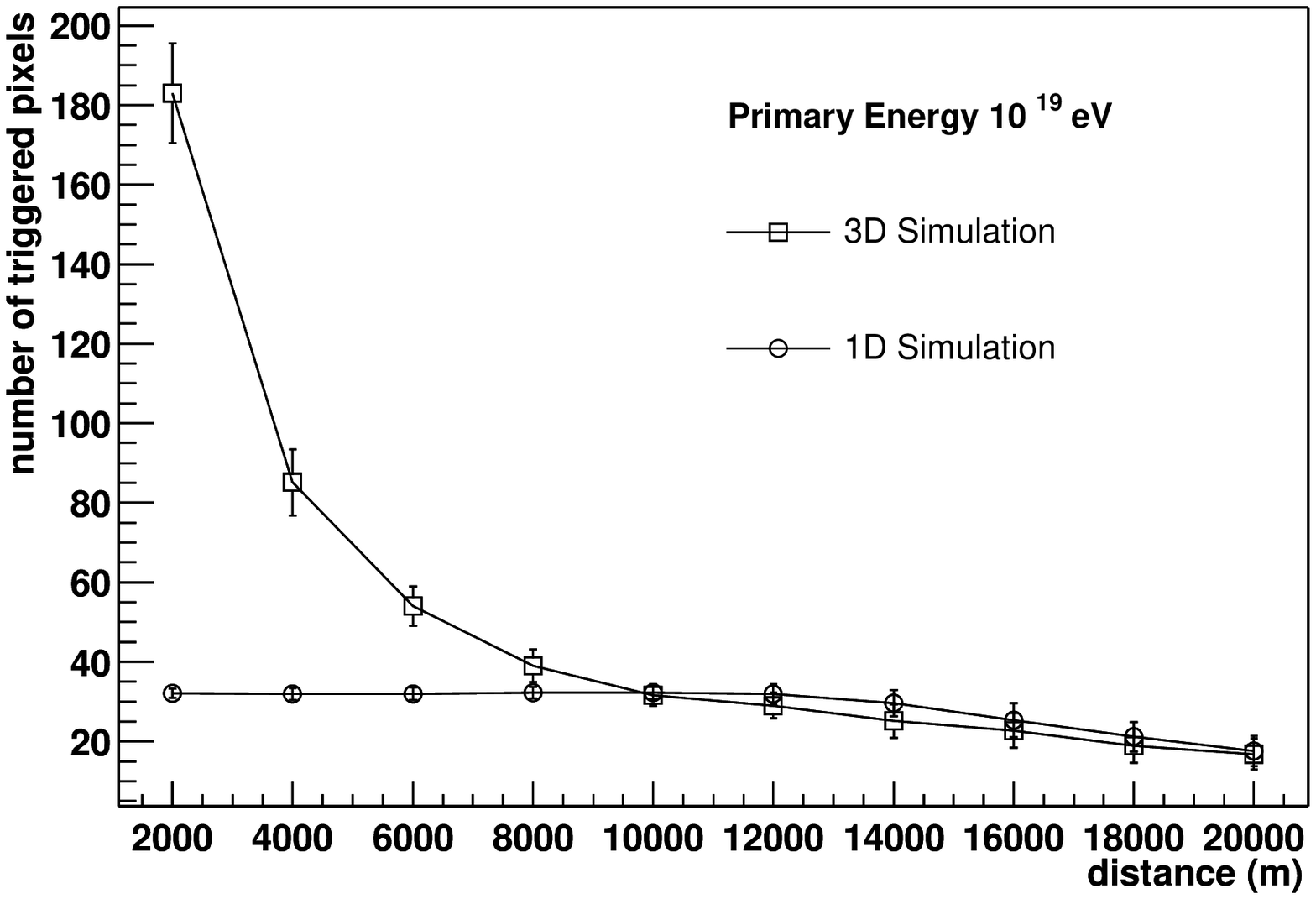}}
\caption{The number of triggered pixels as function of the distance from the core to
the telescope for one and three-dimensional simulations of shower with
primary energy of $10^{19} \ eV$.}
\label{fig:npixdist}
\end{figure}

Figure \ref{fig:cam} shows an example of the number of pixels
triggered in the camera by one-dimensional and three-dimensional
methods for a shower falling $4 \ km$ far from the telescope.
Adopting a simple trigger condition given by those pixels with signal
4$\sigma$ above the average of the noise in that event, we are able to
measure the number of pixels triggered in the Auger telescopes as a
function of the distance of the shower core for one-dimensional
showers and for three-dimensional showers. Figure \ref{fig:npixdist}
shows this comparison. The number of pixels triggered by the shower
simulated with the one and three-dimensional approaches are, within
the statistical variation, the same for distances greater than $8 \
km$.

Another parameter that can give us the sensitivity of the detector to
the lateral distribution of showers is the radius of a circle ($\zeta$
measured in degrees) on the photomultiplier camera which maximizes the
signal to noise ratio, $S/N$. The distribution of the triggered pixels
in a camera allows us to determine the main track of the shower.  The
method to determine $\zeta$ consists in searching the radius of the
circle centered on the track along the entire path which maximizes
$S/N$.

\begin{figure}[p]
\centerline{\includegraphics[width=13cm]{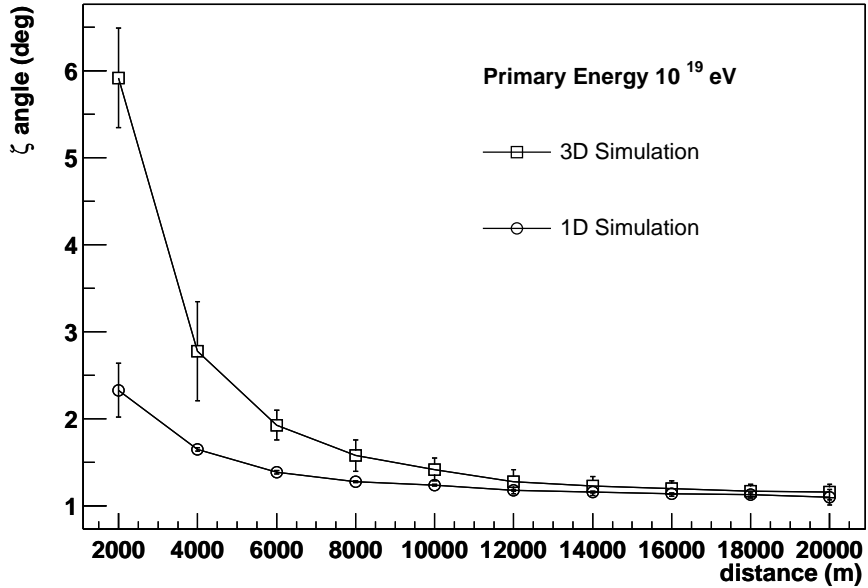}}
\caption{The $\zeta$ angle as function of the distance from the core to
the telescope for one and three-dimensional simulations of showers  with
primary energy of $10^{19} \ eV$.}
\label{fig:zetadist}
\end{figure}

Figure \ref{fig:zetadist} shows the curve of $\zeta$ obtained for one
and three-dimensional simulations as a function of the distance
between the telescope and the core position of the shower. The $\zeta$
angle increases for small distances for the one-dimensional
simulations because of the scattered \v{C}erenkov light included in
the simulations.

Figure \ref{fig:zetadist} shows that for core distances closer than $8
\ km$ of the telescope the lateral distribution of the particles in
the shower produces a measurable spread of the signal in the
photomultiplier camera. The $\zeta$ angle increases quickly as the
core gets closer to the telescope.

\subsection{Energy Dependency}

The capability of a fluorescence telescope to distinguish between
one-dimensional and three-dimensional showers certainly depends on the
primary energy of the shower. With increasing energy, the lateral
spread of the particles gets larger in comparison with one-dimensional
approximations and it should be easier to measure the width of the
signal.

In order to study the energy dependency we simulated 50 vertical
proton showers with primary energy $10^{18} \ eV$ and 10 vertical
proton showers with primary energy $10^{20} \ eV$. The number of
events was reduced in comparison to the previous analysis because the
errors had shown to be small. We used the same thinning factor set to
$10^{-6}$.

\begin{figure}[p]
\centerline{\includegraphics[width=13cm]{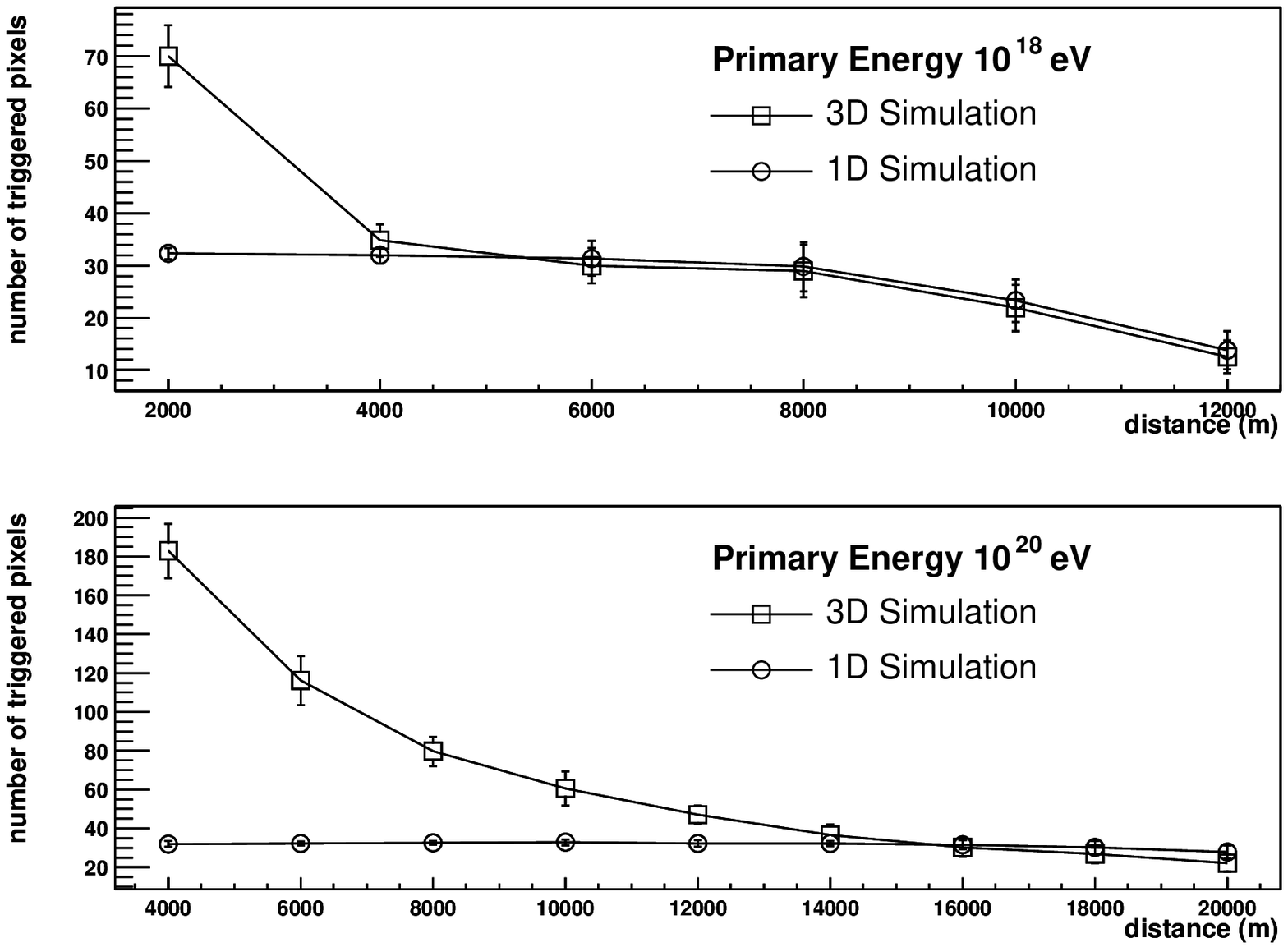}}
\caption{The number of triggered pixels as function of the distance from the core to
the telescope for one and three-dimensional simulations of shower with
primary energy of $10^{18} \ eV$ and $10^{20} \ eV$.}
\label{fig:tudo}
\end{figure}

Figure \ref{fig:tudo} shows the number of pixels triggered as a
function of the distance from the shower axis. We show the comparison
between one-dimensional and three-dimensional simulations. Similar
plots for the $\zeta$ angle have been done and show the same
results. We have limited the distance from the axis to the shower in
the region we expected to find the separation between the one and
three dimensional approaches. Based on figures \ref{fig:npixdist} and
\ref{fig:tudo} we are able to determine the maximum distance that a
fluorescence telescope may be able to measure the lateral distribution
of a shower as a function of energy as being $4 \ km$ for $10^{18} \
eV$, $8 \ km$ for $10^{19} \ eV$ and $14 \ km$ for $10^{20} \ eV$
showers.

\subsection{Evaluating the influence of the \emph{thinning}}

The distribution of weights given as a function of the lateral
distribution of the shower is not constant and depends on the particle
type.  Therefore, the unthinning procedure for the production of
fluorescence photons proposed in section \ref{sec:simulation} should
be tested in order to guarantee that no spurious fluctuations have
been introduced in the determination of the lateral size of the shower
seen by a fluorescence telescope.

\begin{figure}[p]
  \centerline{\includegraphics[width=13cm]{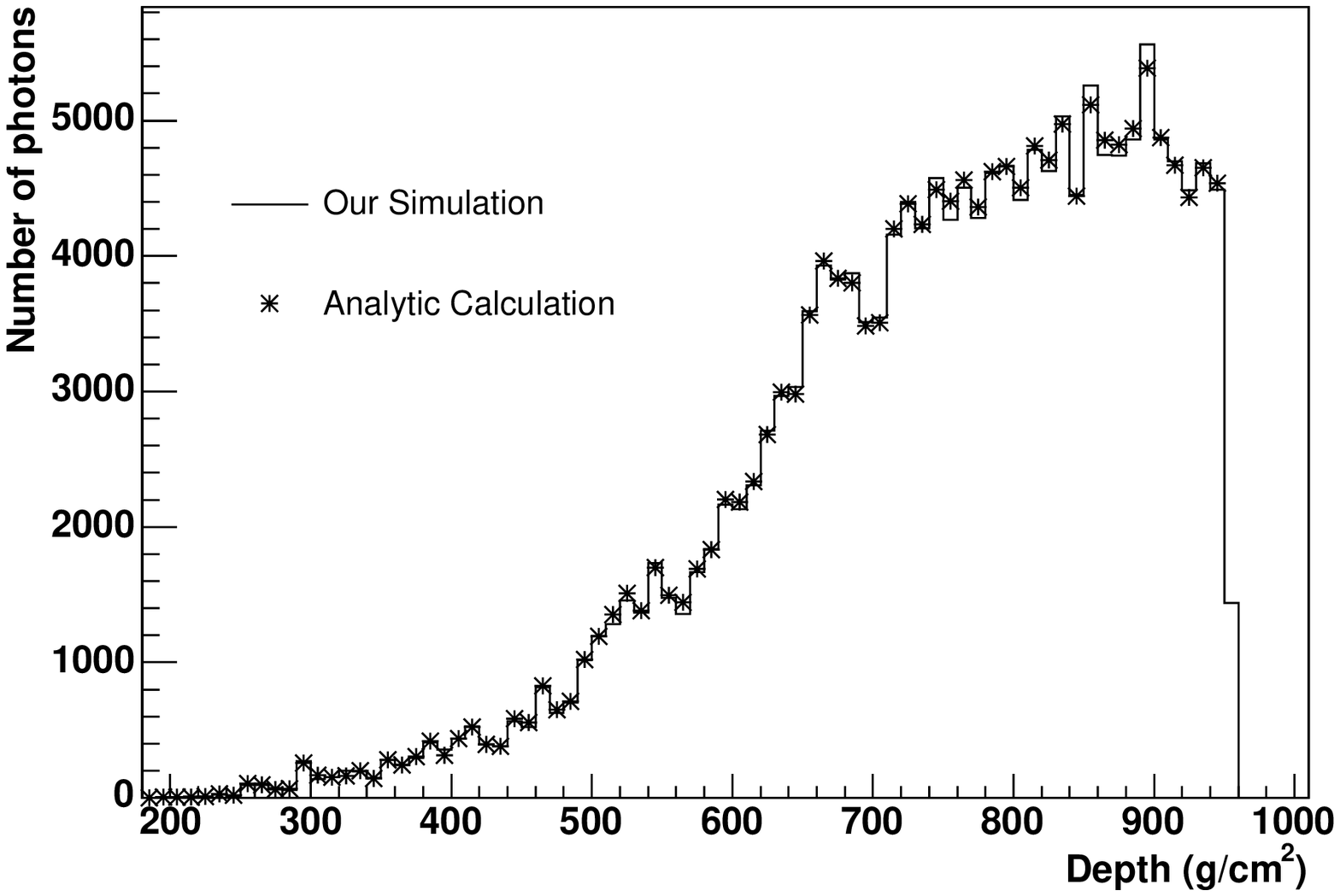}}
\caption{Comparison between the number of photons as a function of
  atmospheric depth calculated from the
  longitudinal energy deposited in the atmosphere and simulated by our
  program. The thinning factor as set to $10^{-4}$.}
\label{fig:perfil:thin}
\end{figure}

Figure \ref{fig:perfil:thin} shows that the unthinning procedure
implemented here is not able to smooth the fluctuations in the
longitudinal profiles. In this figure, we show the number of photons
detected by the telescope as a function of depth for one vertical
shower initiated by an electron with energy $10^{18} \ eV$. The
thinning factor was set to $10^{-4}$. The telescope is $40 \ km$ away
from the axis of the shower and has $10 \ m$ radius. We have set the
same configuration of the shower presented in figure \ref{fig:perfil}
except for the thinning factor. The analytic calculations were carried
out in the same way as explained in section \ref{sec:certification}.

The comparison between figure \ref{fig:perfil} (Thinning Factor
$10^{-6}$) and figure \ref{fig:perfil:thin} (Thinning Factor
$10^{-4}$) illustrates the fluctuations caused by a strong shower
thinning. Nevertheless, figure \ref{fig:perfil:thin} shows our
simulation produces the expected number of photons independently of
that factor.

However, in this article we would like to guarantee that the results
shown in section \ref{sec:application} are independent of the thinning
factor. Therefore, we have also used the number of pixels triggered in
the camera and the $\zeta$ angle to verify the thinning influence on
the final measurement of the lateral spread of the showers. We have
simulated ten vertical proton initiated showers with energy $10^{19} \
eV$ for different thinning factors ($10^{-4}$,$10^{-5}$, $10^{-6}$ and
$10^{-7}$) through the Auger telescopes and determined the
distribution of the number of pixels triggered and $\zeta$ as
described above.

Table \ref{tab:thinn} shows the average number of pixels triggered and
$\zeta$ angle for each thinning factor for telescopes at $4 \ km$ and
$8 \ km$ away from the shower axis. As can be seen, the unthinning
algorithm does not introduce any detectable fluctuation in the shower
lateral size measured by Auger type telescopes for the thinning factor
ranging from $10^{-4}$ to $10^{-7}$.

\section{Conclusions}

We have presented here a new approach to simulate fluorescence photons
in an atmospheric air shower. A three-dimensional and fully hybrid
(particles+photons) simulation tool was developed under the CORSIKA
framework. In our routines, fluorescence photons were produced for all
charged particles in a shower and for electrons and positrons with
energy below the CORSIKA cut-off.

This program is going to be useful for the new generation of detectors
(Auger, EUSO, Telescope Array and OWL) due to the improvement in the
technology used by them.  As shown here, for the new observatories the
shower can not be considered one-dimensional in all cases and
therefore a three-dimensional approach should be used to optimize the
construction and in the elaboration of reconstruction methods.

In this paper, the number of photons produced along the longitudinal
development of the shower and the arrival time of the photons in the
detector determined by our program have been compared to analytical
calculations based on the energy deposited profile given by CORSIKA.

A good agreement can be seen in figures \ref{fig:perfil} and
\ref{fig:time}.  The famous $320 \ EeV$ shower detected by the Fly's
Eye experiment was also simulated according to the geometry given in
\cite{bib:fly51joule}. The consistency of our simulation can be seen
as a test of the approach proposed here from one hand and as
confirmation of the profile reconstructed by them from the other hand.

Our code was approved in several comparisons of simulated and real
data. The arrival time, the longitudinal profile and the energy
determined by our simulation show a very good agreement to what has
been published so far.

The fluctuation due to thinning algorithm introduced in section
\ref{sec:simulation} has been tested in respect to its influence on
the lateral distribution of particles. No effect of the thinning was
found in a range from $10^{-4}$ to $10^{-7}$ of the thinning factor
when detectors similar to the Auger telescopes configuration are used.

Applications of the code were performed. We determined the sensitivity
of a fluorescence telescope of the kind used by the Auger
collaboration to the lateral distribution of the particles in a
shower. Such a telescope could distinguish a real shower from a one
dimensional approximation for distances of the core closer than $4 \
km$ for $10^{18} \ eV$ showers, $8 \ km$ for $10^{19} \ eV$ showers
and $14 \ km$ for $10^{20} \ eV$ showers.

These results represent a limit on the possibility to measure
the lateral distribution of particles with a fluorescence telescope
based on time information as suggested by
\cite{bib:sommers_size}. However, a great number of showers with
energy beyond $10^{19} \ eV$ with core closer $8 \ km$ can be detected
and reconstructed by the present and future fluorescence
telescopes. For this subset of events we suggest a detailed analysis
of the data in order to possibly measure the lateral distribution of
the particles.

\section{Acknowledgments}

The authors would like to thank Dieter Heck for very fruitful
discussions during the development of this code. This work was
supported by the Brazilian population via the science foundations
FAPESP and CAPES to which we are grateful. The calculations were done
at Campinas using the computational facilities funded by FAPESP.  We
are also very grateful to Carlos Escobar for reading the manuscript
and for useful discussions.

\bibliographystyle{elsart-num}

\bibliography{corsika-astro}

\newpage

\begin{table}[t]
\begin{center}
\begin{tabular}{|c|c|c|c|c|} \hline
Thinn. Factor    &  \multicolumn{2}{|c|}{$\zeta$ Angle (deg)} &
           \multicolumn{2}{|c|}{N Pixels Trigg.} \\ \hline
          &    $4 \ km$    &    $8 \ km$   &  $4 \  km$  & $8 \ km$ \\ \hline
$10^{-4}$ &  2.4$\pm$0.87 & 1.5$\pm$0.1 & 87.7$\pm$3.1 &
           35.3$\pm$3.9 \\ \hline
$10^{-5}$ &  2.5$\pm$0.1 & 1.4$\pm$0.1 & 78.6$\pm$6.4  &
           35.6$\pm$2.6 \\ \hline 
$10^{-6}$ & 2.4$\pm$0.06 & 1.4$\pm$0.1 & 82.0$\pm$4.5 &
           35.5$\pm$2.3 \\ \hline
$10^{-7}$ & 2.4$\pm$0.1 & 1.4$\pm$0.1 & 76.3$\pm$4.5 &
           36.0$\pm$2.4 \\ \hline
\end{tabular}
\caption{Number of pixels triggered and $\zeta$ angle for different
  thinning factors and two telescopes at $4 \ km$ and $8 \ km$ away from the
  shower axis. The values are the average and standard deviation for a
  sample of ten vertical proton induced showers with energy $10^{19} \
  eV$.} 
\label{tab:thinn}
\end{center}
\end{table}

\end{document}